# Conducting LaAlO$_3$/SrTiO$_3$ heterointerfaces on atomically-flat substrates prepared by deionized-water

J. G. Connell,[1] J. Nichols,[1] J. H. Gruenewald,[1] D.-W. Kim,[2] and S. S. A. Seo[1]

[1] *Department of Physics and Astronomy, University of Kentucky, Lexington, KY 40506, USA*

[2] *Department of Physics, Ewha Womans University, Seoul 120-750, Korea*

We have investigated how the recently-developed water-leaching method for atomically-flat SrTiO$_3$ (STO) substrates affects the transport properties of LaAlO$_3$ (LAO) and STO heterointerfaces. Using pulsed laser deposition at identical growth conditions, we have synthesized epitaxial LAO thin-films on two different STO substrates, which are prepared by water-leaching and buffered hydrofluoric acid (BHF) etching methods. The structural, transport, and optical properties of LAO/STO heterostructures grown on water-leached substrates show the same high-quality as the samples grown on BHF-etched substrates. These results indicate that the water-leaching method can be used to grow complex oxide heterostructures with atomically well-defined heterointerfaces without safety concerns.

* Correspondence and requests for materials should be addressed to S.S.A.S. (email: a.seo@uky.edu)

# Introduction

Preparation of atomically-flat surfaces of substrates is an important step to successfully fabricate well-characterized epitaxial thin-films and heterointerfaces. For example, the atomically-flat $TiO_2$-terminated $SrTiO_3$ surface[1–4] is the key for creating the high-mobility two-dimensional election gas (2DEG) at $LaAlO_3/SrTiO_3$ (LAO/STO) heterointerfaces, which show intriguing multichannel conduction,[5–8] interfacial superconductivity,[9] ferromagnetism,[10,11] and for developing electronic devices and sensors.[12–14] The atomically-flat surfaces of $SrTiO_3$ (STO) single crystal substrates are usually achieved by an acid-based chemical etching procedure[15–17] followed by thermal-annealing. For example, buffered-hydrofluoric acid (BHF), which is used in silicon semiconductor research and industry for removing $SiO_2$, has been used widely for making atomically-flat STO substrates.[15–20] Recently, we have shown that a *non-acidic* deionized-water-leaching method is as effective at generating single-terminated atomically-flat STO substrates.[21] The water-leaching method can remove effectively SrO, which is a hydrophilic compound,[19–27] from the STO surface. Since 6-14% of fluorine impurities can be doped into the STO surface during the BHF-etching procedure,[24] water-leaching eliminates not only the safety concerns of acidic etchants but also possible impurity doping on the surface.

In this article, we report that the water-leaching method creates heterointerfaces that have the same high-quality as those generated through the BHF-etching method. We have investigated the LAO/STO 2DEG as a representative model system requiring atomically well-defined STO substrates. By *simultaneously* depositing $LaAlO_3$ (LAO) films under the same condition on two STO substrates, which are prepared by the water-leaching and BHF-etching methods, respectively, we have observed that there is no noticeable difference between the two heterointerfaces regarding their structural, transport, and optical properties.

**Methods**

We have synthesized LAO/STO heterointerfaces on atomically-flat surfaces of (100)-oriented STO substrates that are prepared by using either the water-leaching[21] or BHF-etching methods.[15] All substrates (purchased from CrysTec GmbH) are sliced into two pieces (5 × 2 × 1 mm$^3$), are annealed at 1000 °C in ambient conditions for 1 hour, which forms a dual-terminated step-and-terrace structure, and each piece is leached (etched) for 30 s in deionized-water (BHF), respectively. Substrates are again annealed at 1000 °C for 2 hours, which effectively forms atomically-flat single-terminated surfaces, reducing the overall surface roughness. The final step of substrate preparation is to once again leach in deionized-water or etch in BHF as before in order to eliminate possible strontium oxides or strontium hydroxides segregated on the surface.[21] Atomic Force Microscopy (Park XE-70) is employed to ensure the formation of single-terminated atomically-flat substrates before deposition and to confirm film surface quality after deposition. Epitaxial LAO thin-films of various thickness (5 – 60 unit-cells) are deposited on the STO substrates using pulsed laser deposition with a laser fluence (KrF excimer, $\lambda$ = 248 nm) of 1.6 J/cm$^2$, a substrate temperature of 700 °C, and pO$_2$ of 10$^{-6}$ Torr. *In situ* reflection high energy electron diffraction (RHEED) is utilized to monitor the number of unit cells of LAO deposited. The grown samples are cooled naturally for 2 hours to room temperature at a higher oxygen partial pressure (10 mTorr) so that the films have proper oxygen stoichiometry. There is no clear systematic thickness dependence of LAO thin-films on their transport properties, as reported previously.[28] Thus, here we focus our discussion on the results obtained from the 5, 25, and 30 unit-cell LAO samples. Structural quality of the films is characterized using X-ray diffractometry (Bruker D8 Advance). Optical transmission spectra is taken at room temperature using a Fourier-transform infrared spectrometer (FT-IR) (for spectra regions between 50 meV and 0.6 eV) and a grating-type spectrophotometer (for spectra regions between 0.5 and 6 eV). Transport properties are measured using a Physical Property Measurement System (Quantum Design) with conventional four-probe and Hall geometries. Hall measurements are taken at various temperatures at a maximum magnetic field of 9 T. Electrical contacts are made using aluminum wire attached with indium solder, which gives access to the 2DEG present at the heterointerface.

**Results and Discussion**

LAO thin-films deposited on water-leached STO substrates show the same film quality as BHF-etched substrates. Figure 1 (a) depicts a 3 × 3 μm$^2$ atomic force microscopy (AFM) topography scan of a water-leached STO substrate with respective line profile below. As indicated in the line profile, the substrate has a step height of 3.9 Å, which is the lattice constant of cubic STO. Fig. 1 (b) displays the same sample as in (a) after deposition of a 30 unit-cell LAO film. Both images show single-terminated atomically-flat step terraces before and after deposition. Fig. 1 (c) displays the RHEED intensity oscillations for the 5 unit-cell thick LAO film deposited on the water-leached substrate. The insets show the RHEED patterns at the beginning and end of film deposition, which, other than a change in intensity, do not display any noticeable changes. The high quality of the LAO films is confirmed further by the X-ray $\theta$-$2\theta$ scans, as shown in Fig. 1 (d) for the 30 unit-cell thick films. The peak position of the (220)-LAO plane does not depend on substrate preparation method. The X-ray reciprocal space maps near the (114)-STO reflection show that both LAO thin-films exhibit coherent in-plane tensile strain with no evidence of strain relaxation, as shown in Fig. 1 (e) for water-leached and 1 (f) for BHF-etched samples.

The optical transmission spectra of both heterointerfaces show little difference in the range of 0.2 – 3.2 eV, demonstrating that their optical properties and electronic structures are quite similar regardless of substrate preparation method. Figure 2 illustrates the optical transmittance spectra of the 25 unit-cell LAO/STO grown on water-leached and BHF-etched substrates. Both spectra demonstrate clear Drude absorption due to conducting carriers, i.e. the decrease of optical transmittance, below about 1.5 eV. These transmittance spectra are consistent with the optical properties of LAO/STO heterointerfaces, reported in Ref 5. The three dip structures near 1.7, 2.4, and 2.9 eV are commonly observed in LAO/STO heterostructures and reduced STO crystals. The absorption at 1.7 eV increases as STO crystals are reduced, hence it is related to the oxygen vacancy level.[29] The dip structures at 2.4 eV and 2.9 eV are observed regardless of free carrier concentration, and they may originate from the excitation of electrons trapped by oxygen vacancies, i.e. $F_1$ centers.[30]

The sheet resistance of both heterointerfaces has similar behavior down to low temperatures, regardless of substrate preparation method. Figure 3 shows the sheet resistance as a function of temperature for the LAO/STO heterointerfaces for the 5 unit-cell and 30 unit-cell LAO layers. The sheet resistance of the same LAO thickness is qualitatively identical despite the use of two different methods of substrate preparation. It is noteworthy that the 30 unit-cell LAO/STO samples display metal-insulator transitions at around 40 K while the 5 unit-cell LAO/STO samples are overall metallic. This behavior has been reported previously: the resistivity of LAO/STO heterointerfaces with thicker LAO layers can be larger than that of thinner samples, which may be due to structural reconstructions at the LAO/STO interface.[28]

The heterointerfaces also have comparable carrier concentrations and mobilities. The results of the Hall measurements for the metallic 5 unit-cell LAO/STO heterointerfaces are displayed in Figures 4 (a, b). Sheet carrier concentration ($n_s$) and mobility ($\mu$) of the heterointerfaces prepared on the two kinds of STO substrates are similar regardless of preparation method. The values of $n_s$ and $\mu$ compare well to those of other conducting LAO/STO 2DEG's where similar deposition conditions were used.[1,3,4,28] Further, room temperature $n_s$ exceeding or near $10^{13}$-$10^{14}$ cm$^{-2}$ are observed in most LAO/STO 2DEG's when the pO$_2$ of deposition is below $10^{-5}$ Torr.[1,3,4,28] Thus, as in most low-pO$_2$ LAO/STO heterointerfaces, oxygen vacancies play a role in the heterointerfacial conductivity. It is noteworthy that below 100 K we observe the non-linear Hall effect due to multi-channel electron conduction as shown in the inset of Fig. 4 (b) as has been seen previously.[6–8] This effect can be fitted by a two-band model, assuming the same sign for the charge carriers.[31] Thus, we can write the Hall coefficient, $R_H = \frac{R_{XY}}{B}$, as $R_H = \left(\frac{1}{e}\right)\frac{n_1\mu_1^2+n_2\mu_2^2+(n_1+n_2)\mu_1^2\mu_2^2 B^2}{(n_1\mu_1+n_2\mu_2)^2+(n_1+n_2)^2\mu_1^2\mu_2^2 B^2}$.[31–33] We can rewrite this equation of four unknown parameters as an equation of two unknown parameters, $R_H = \frac{R_0+R_\infty \mu_*^2 B^2}{1+\mu_*^2 B^2}$, where $\mu_*$ and $R_\infty$ are fitting parameters with $R_0$ being $R_H(B=0)$. Using the zero field resistivity, $R_{XX} = (en_1\mu_1 + en_2\mu_2)^{-1}$, we can find the low-density-high-mobility (LDHM) ($n_2$ and $\mu_2$) and high-density-low-mobility (HDLM) ($n_1$ and $\mu_1$) carriers using $A = \frac{1}{2}\left(\frac{R_0}{R_{XX}}+\mu_*\right)$, $\mu_1 = A + (A^2 - \frac{\mu_* R_\infty}{R_{XX}})^{1/2}$, $\mu_2 = A - (A^2 - \frac{\mu_* R_\infty}{R_{XX}})^{1/2}$, $C = \frac{\mu_1(\mu_*-\mu_2)}{\mu_2(\mu_1-\mu_*)}$, $n_1 = \frac{1}{eR_\infty(1+C)}$, and $n_2 = \frac{C}{eR_\infty(1+C)}$.[33–35] The model fits at 50 and 2 K are shown by the black lines in the inset of Fig. 4 (b). As stated above, neither the LDHM nor the HDLM display any differences based on substrate preparation. According to Ref. 24, the BHF-etching method might result in a few percent of fluorine doping into STO, which can provide $4 \times 10^{13}$ cm$^{-2}$ to $1 \times 10^{14}$ cm$^{-2}$ extra carriers.[24] Fig. 4 (a), however, shows that $n_s$ for both samples is very similar in the whole measurement temperature range. As for changes in $\mu$, Ref. 24 also suggests that the fluorine atoms, acting as impurity sites, would increase the scattering rate, thereby reducing the overall $\mu$ of any heterointerface. However, the $\mu$ of the two kinds of 5 unit-cell 2DEG samples shows little or no difference, as displayed in Fig. 4 (b). Thus, fluorine doping does not appear to alter the electronic properties of oxygen-deficient conducting LAO/STO heterointerfaces.

**Conclusion**

LAO/STO heterointerfaces grown on water-leached and BHF-etched STO substrates show similar structural, optical, and electronic properties. Based on these results, the water-leaching method produces not only atomically-flat single-terminated surfaces of STO but also high-quality heterointerfaces of complex oxides. Recently, various oxide heterointerfaces grown on STO substrates such as $LaTiO_3/SrTiO_3$,[31,36,37] $LaVO_3/SrTiO_3$,[33,38] $LaMnO_3/SrTiO_3$,[39] $GdTiO_3/SrTiO_3$,[40] $NdAlO_3/SrTiO_3$,[41] and $NdGaO_3/SrTiO_3$[41,42] have demonstrated intriguing electronic reconstructions, interfacial superconductivity, and magnetic ordering. Hence, the use of the water-leaching method promotes research on future oxide electronics by providing a safe way to prepare atomically-flat complex-oxide substrates.

**References**

1. Ohtomo, A. & Hwang, H. Y. A high-mobility electron gas at the $LaAlO_3/SrTiO_3$ heterointerface. *Nature* **427**, 423–426 (2004).
2. Herranz, G. *et al.* High Mobility in $LaAlO_3/SrTiO_3$ Heterostructures: Origin, Dimensionality, and Perspectives. *Phys. Rev. Lett.* **98**, 216803 (2007).
3. Siemons, W. *et al.* Origin of Charge Density at $LaAlO_3$ on $SrTiO_3$ Heterointerfaces: Possibility of Intrinsic Doping. *Phys. Rev. Lett.* **98**, 196802 (2007).
4. Kalabukhov, A. *et al.* Effect of oxygen vacancies in the $SrTiO_3$ substrate on the electrical properties of the $LaAlO_3/SrTiO_3$ interface. *Phys. Rev. B* **75**, 121404 (2007).
5. Seo, S. S. A. *et al.* Multiple conducting carriers generated in $LaAlO_3/SrTiO_3$ heterostructures. *Appl. Phys. Lett.* **95**, 082107 (2009).
6. Ben Shalom, M., Ron, A., Palevski, A. & Dagan, Y. Shubnikov–De Haas Oscillations in $SrTiO_3/LaAlO_3$ Interface. *Phys. Rev. Lett.* **105**, 206401 (2010).

7. Lerer, S., Ben Shalom, M., Deutscher, G. & Dagan, Y. Low-temperature dependence of the thermomagnetic transport properties of the SrTiO$_3$/LaAlO$_3$ interface. *Phys. Rev. B* **84**, 075423 (2011).

8. Fête, A. *et al.* Growth-induced electron mobility enhancement at the LaAlO$_3$/SrTiO$_3$ interface. *Appl. Phys. Lett.* **106**, 051604 (2015).

9. Reyren, N. *et al.* Superconducting Interfaces Between Insulating Oxides. *Science* **317**, 1196–1199 (2007).

10. Bert, J. A. *et al.* Direct imaging of the coexistence of ferromagnetism and superconductivity at the LaAlO$_3$/SrTiO$_3$ interface. *Nat. Phys.* **7**, 767–771 (2011).

11. Dikin, D. A. *et al.* Coexistence of Superconductivity and Ferromagnetism in Two Dimensions. *Phys. Rev. Lett.* **107**, 56802 (2011).

12. Xie, Y., Bell, C., Hikita, Y., Harashima, S. & Hwang, H. Y. Enhancing electron mobility at the LaAlO$_3$/SrTiO$_3$ interface by surface control. *Adv. Mater.* **25**, 4735–4738 (2013).

13. Kim, H. *et al.* Influence of Gas Ambient on Charge Writing at the LaAlO$_3$/SrTiO$_3$ Heterointerface. *ACS Appl. Mater. Interfaces* **6**, 14037–14042 (2014).

14. Kim, S. K. *et al.* Electric-field-induced Shift in the Threshold Voltage in LaAlO$_3$/SrTiO$_3$ Heterostructures. *Sci. Rep.* **5**, 8023 (2015).

15. Kawasaki, M. *et al.* Atomic Control of the SrTiO$_3$ Crystal Surface. *Science* **266**, 1540–1542 (1994).

16. Kareev, M. *et al.* Atomic control and characterization of surface defect states of TiO$_2$ terminated SrTiO$_3$ single crystals. *Appl. Phys. Lett.* **93**, 061909 (2008).

17. Biswas, A. *et al.* Universal Ti-rich termination of atomically flat SrTiO$_3$ (001), (110), and (111) surfaces. *Appl. Phys. Lett.* **98**, 051904 (2011).

18. Ohnishi, T. *et al.* Preparation of thermally stable TiO$_2$-terminated SrTiO$_3$(100) substrate surfaces. *Appl. Phys. Lett.* **85**, 272-274 (2004).

19. Tench, D. M. & Raleigh, D. O. Electrochemical Processes on Strontium Titanate Electrodes. *Natl. Bur. Stand. Spec. Publ.* **455**, 229-240 (1976).

20. Koster, G., Kropman, B. L., Rijnders, G. J. H. M., Blank, D. H. A. & Rogalla, H. Quasi-ideal strontium titanate crystal surfaces through formation of strontium hydroxide. *Appl. Phys. Lett.* **73**, 2920-2922 (1998).

21. Connell, J. G., Isaac, B. J., Ekanayake, G. B., Strachan, D. R. & Seo, S. S. A. Preparation of atomically flat SrTiO$_3$ surfaces using a deionized-water leaching and thermal annealing procedure. *Appl. Phys. Lett.* **100**, 215607 (2012).

22. Lide, D. R. *CRC Handbook of Chemistry and Physics*. (CRC Press, 2004).

23. Karthäuser, S., Speier, W. & Szot, K. Verfahren zur Herstellung einer B-terminierten Oberfläche auf Perowskit-Einkristallen. Germany Patent 102004019690 A1 filed 20 Apr. 2004, and issued 10 Nov. 2005.

24. Chambers, S. A., Droubay, T. C., Capan, C. & Sun, G. Y. Unintentional F doping of

SrTiO$_3$(001) etched in HF acid-structure and electronic properties. *Surf. Sci.* **606**, 554–558 (2012).

25. Boschker, J. E. & Tybell, T. Qualitative determination of surface roughness by in situ reflection high energy electron diffraction. *Appl. Phys. Lett.* **100**, 151604 (2012).

26. Hatch, R. C. *et al.* Surface electronic structure for various surface preparations of Nb-doped SrTiO$_3$ (001). *J. Appl. Phys.* **114**, 103710 (2013).

27. Hatch, R. C., Choi, M., Posadas, A. B. & Demkov, A. A. Comparison of acid- and non-acid-based surface preparations of Nb-doped SrTiO$_3$ (001). *J. Vac. Sci. Technol. B* **33**, 061204 (2015).

28. Bell, C., Harashima, S., Hikita, Y. & Hwang, H. Y. Thickness dependence of the mobility at the LaAlO$_3$/SrTiO$_3$ interface. *Appl. Phys. Lett.* **94**, 222111 (2009).

29. Lee, C., Destry, J. & Brebner, J. L. Optical absorption and transport in semiconducting SrTiO$_3$. *Phys. Rev. B* **11**, 2299–2310 (1975).

30. Baer, W. S. Free-carrier absorption in reduced SrTiO$_3$. *Phys. Rev.* **144**, 734–738 (1966).

31. Kim, J. S. *et al.* Nonlinear Hall effect and multichannel conduction in LaTiO$_3$/SrTiO$_3$ superlattices. *Phys. Rev. B* **82**, 201407 (2010).

32. Ziman, J. M. *Principles of the Theory of Solids*. (Cambridge University Press, 1964).

33. Rotella, H. *et al.* Two components for one resistivity in LaVO$_3$/SrTiO$_3$ heterostructure. *J. Phys. Condens. Matter* **27**, 95603 (2015).

34. Arushanov, E. K. & Chuiko, G. P. The Magnetic Field Dependence of Kinetic Coefficients of Cadmium Arsenide Single Crystals. *Phys. Status Solidi* **17**, K135-K138 (1973).

35. Laiho, R. *et al.* Hall effect and band structure of p-CdSb in strong magnetic field. *Semicond. Sci. Technol.* **19**, 602–609 (2004).

36. Ohtomo, A., Muller, D. A., Grazul, J. L. & Hwang, H. Y. Artificial charge-modulationin atomic-scale perovskite titanate superlattices. *Nature* **419**, 378–380 (2002).

37. Seo, S. S. A. *et al.* Optical Study of the Free-Carrier Response of LaTiO$_3$/SrTiO$_3$ Superlattices. *Phys. Rev. Lett.* **99**, 266801 (2007).

38. Hotta, Y., Susaki, T. & Hwang, H. Y. Polar Discontinuity Doping of the LaVO$_3$/SrTiO$_3$ Interface. *Phys. Rev. Lett.* **99**, 236805 (2007).

39. Choi, W. S. *et al.* Charge states and magnetic ordering in LaMnO$_3$/SrTiO$_3$ superlattices. *Phys. Rev. B* **83**, 195113 (2011).

40. Moetakef, P. *et al.* Electrostatic carrier doping of GdTiO$_3$/SrTiO$_3$ interfaces. *Appl. Phys. Lett.* **99**, 232116 (2011).

41. Annadi, A. *et al.* Electronic correlation and strain effects at the interfaces between polar and nonpolar complex oxides. *Phys. Rev. B.* **86**, 085450 (2012).

42. Gunkel, F. *et al.* Stoichiometry dependence and thermal stability of conducting

NdGaO$_3$/SrTiO$_3$ heterointerfaces. *Appl. Phys. Lett.* **102**, 071601 (2013).

## Acknowledgements

We acknowledge the support of National Science Foundation grant DMR-1454200 for sample synthesis, transport measurements, and optical spectroscopy in addition to a grant from the Kentucky Science and Engineering Foundation as per Grant Agreement #KSEF-148-502-14-328 with the Kentucky Science and Technology Corporation. D.W.K. was supported by the National Research Foundation of Korea Grant (No. 2015001948).

## Author Contributions

S.S.A.S. supervised the project for this manuscript. J.G.C. and S.S.A.S. synthesized the samples. J.G.C conducted the structural, electronic, and optical measurements. J.G.C. and J.H.G. conducted the FT-IR measurements. J.G.C., J.N., D.W.K., and S.S.A.S. analyzed the experimental data. All authors wrote and reviewed the manuscript.

## Additional Information

**Competing financial interests:** The authors declare no competing financial interests.

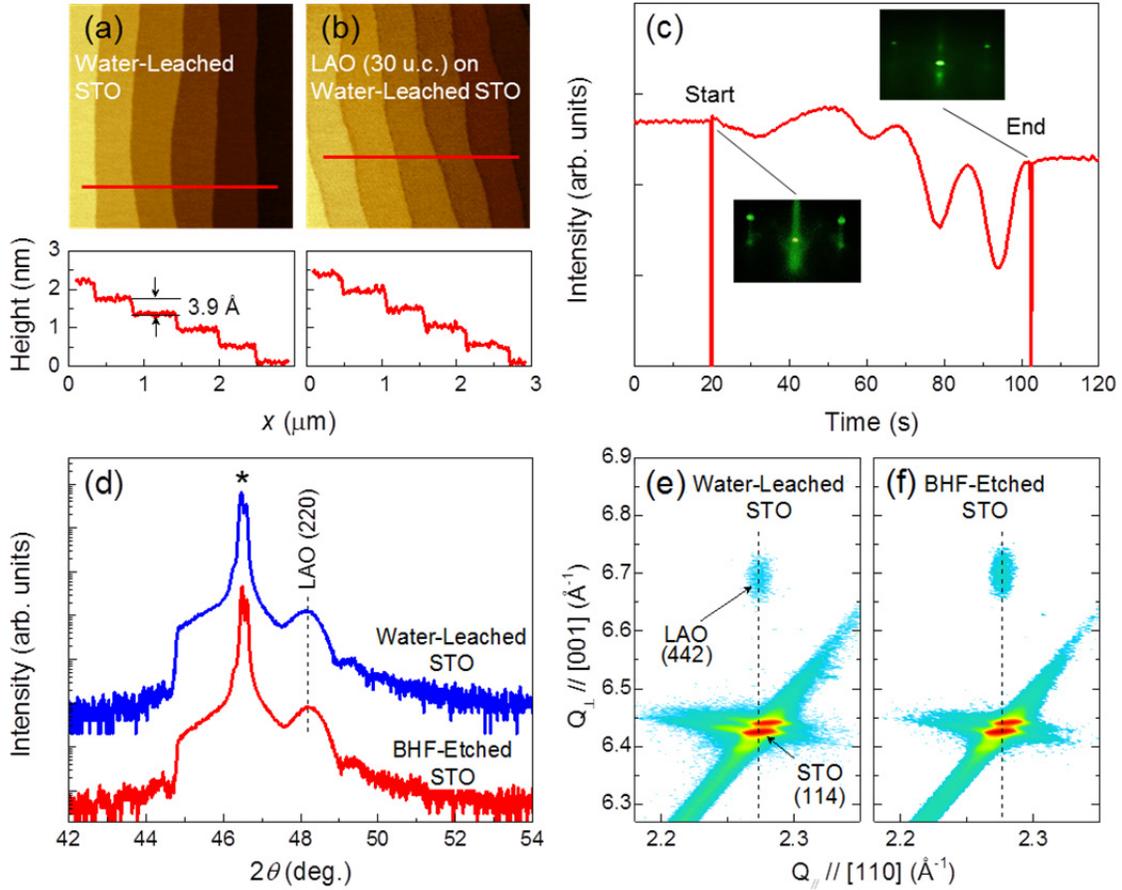

**Figure 1:** Crystal properties of LAO films deposited on water-leached and BHF-etched STO substrates. 3 × 3 μm² AFM topography of a water-leached substrate (a) before and (b) after deposition of a 30 unit-cell LAO thin-film. The red lines correspond to the line profiles shown below each AFM scan. (c) RHEED intensity oscillations of the 5 unit-cell water-leached LAO thin-film deposition on a water-leached STO substrate. The insets show the RHEED pattern at the beginning and end of the thin-film growth. (d) X-ray $\theta$-$2\theta$ scans of the 30 unit-cell water-leached sample (blue) and BHF-etched sample (red). The asterisk (*) indicates the STO substrate (200) reflection. X-ray reciprocal space maps near the STO (114) reflection for the 30 unit-cell LAO thin-films deposited on (e) water-leached and (f) BHF-etched substrates. Note that both LAO thin-films are coherently strained.

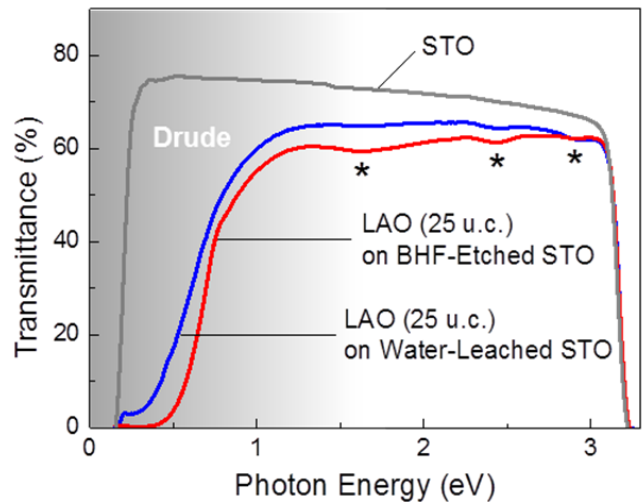

**Figure 2:** Optical transmittance spectra of 25 unit-cell LAO thin-films grown on water-leached (blue) and BHF-etched (red) STO substrates. The STO substrate (grey) is shown for comparison. The asterisks (*) at 1.7, 2.4 and 2.9 eV indicate the optical absorptions due to oxygen vacancies. The shaded region below about 1.5 eV indicates the decrease of optical transmittance due to conducting Drude carriers. Two sudden drops of optical transmittance at 0.2 eV and 3.2 eV are due to the Reststrahlen band and the bandgap energy of STO, respectively.

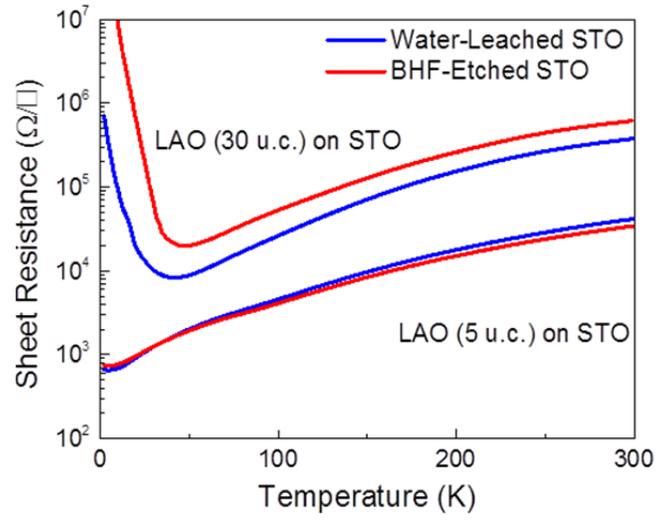

**Figure 3:** Temperature dependence of the sheet resistance of the 5 and 30 unit-cell LAO thin-films grown on water-leached (blue) and BHF-etched (red) STO substrates.

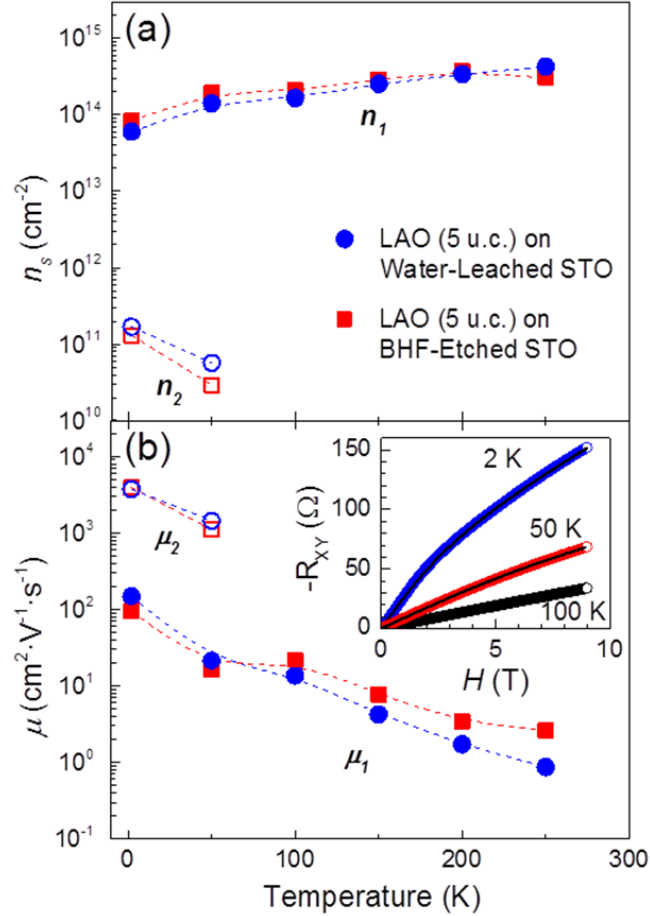

**Figure 4:** Temperature dependence of the (a) sheet carrier concentration and (b) mobility for the 5 unit-cell LAO/STO heterointerfaces. The filled circles (squares) indicate the high-density low-mobility carriers and open circles (squares) indicate the low-density high-mobility carriers for samples grown on the water-leached (BHF-etched) STO substrates. The dotted lines in (a) and (b) are guides for the eye. The inset in (b) shows the Hall resistance as a function of magnetic field at 2, 50, and 100 K for the sample grown on a water-leached STO substrate. The black lines are the two-carrier model fits.